\documentclass[11pt,a4paper]{article}
\pdfoutput=1
\usepackage{jheppub}
\usepackage{graphicx}
\newcommand{\ba}{\begin{eqnarray}}
\newcommand{\ea}{\end{eqnarray}}
\newcommand{\no}{\nonumber}
\newcommand{\be}{\begin{equation}}
\newcommand{\ee}{\end{equation}}
\newcommand{\bea}{\begin{eqnarray}}
\newcommand{\eea}{\end{eqnarray}}

\title{Direct detection of dark matter polarizability
}
\date{\today
}
\author[a]{Grigory Ovanesyan,}
\author[b]{
Luca Vecchi}
\affiliation[a]{Physics Department, University of Massachusetts Amherst, Amherst, MA 01003, USA}
\affiliation[b]{Maryland Center for Fundamental Physics, Department of Physics, University of Maryland,\\ College Park, MD 20742, USA}
\emailAdd{ovanesyan@umass.edu}
\emailAdd{vecchi@umd.edu}
\abstract{
We point out that the direct detection of dark matter via its electro-magnetic polarizability is described by two new nuclear form factors, which are controlled by the 2-nucleon nuclear density. The signature manifests a peculiar dependence on the atomic and mass numbers of the target nuclei, as well as on the momentum transfer, and can differ significantly from experiment to experiment. We also discuss UV completions of our scenario. 
}
\preprint{\vbox{\hbox{ACFI-T14-19}}}
\begin{document}
\maketitle
%

\section{Motivations}

Dark matter (DM) with non-vanishing couplings to ordinary matter may be probed in underground direct-detection experiments. Such couplings can arise from short-range interactions with protons and neutrons, or via weak interactions with photons.

The latter are particularly relevant whenever the DM field $X$ directly couples to messengers that carry electro-weak charges, but couple only very weakly to gluons, the standard model (SM) fermions, and the Higgs boson. In these scenarios, complex DM with spin will generically acquire electro-magnetic dipole moments. These lead to very large direct detection (DD) signatures, and have already been studied by many authors~\cite{Bagnasco:1993st}\cite{Pospelov:2000bq}\cite{Sigurdson:2004zp}\cite{Barger:2010gv}\cite{Banks:2010eh}\cite{Vecchi:2013iza}. 

Here we are interested in the alternative scenarios with self-conjugate $X$ (real scalar, Majorana fermion, real vector, etc.) in which DD is controlled by the DM electro-magnetic polarizability. We define the latter according to 
\ba\label{Ogamma}
\delta{\cal L}\supset\frac{C_\gamma}{\Lambda^3}O_\gamma,~~~~~~~~~~{O}_\gamma=F_{\mu\nu}F^{\mu\nu}\,\overline{X}X,
\ea
with $F^{\mu\nu}$ the photon field strength.~\footnote{At scales relevant to DD experiments, couplings to the intermediate $W^\pm,Z^0$ bosons effectively describe short-range interactions between DM and nucleons.} For definiteness we assume $X$ is a Majorana fermion, but our results apply to self-conjugate DM of any spin. Strictly speaking, there is another $C$ and $P$ invariant operator that contributes to the DM polarizability at leading order in a momentum expansion. In the non-relativistic limit we may write it as $F_{\mu\alpha}F_{\nu\alpha}v^\mu v^\nu\,\overline{X}X$, where $v^\mu$ is the DM 4-velocity. Note that one of the two DM velocities arises from a derivative on the DM field (if the DM is a boson, both of them). Therefore, the latter operator is parametrically suppressed compared to $O_\gamma$ when the scale of the charged mediator inducing (\ref{Ogamma}) is much heavier than the DM mass, but cannot be neglected when the two scales approach each other. The important point for us is that both operators lead, up to small velocity-suppressed corrections, to the same direct detection matrix element, so a distinction between the two is not relevant to our work. In fact, in the limit of small DM velocity $v^\mu\to(1,\overrightarrow{0})$, and the second operator effectively reduces to $F_{0i}F_{0i}\,\overline{X}X$. Analogously, because the photon field in $O_\gamma$ dominantly couples to the zeroth component of the nucleon current, $F_{\mu\nu}F^{\mu\nu}\,\overline{X}X\to F_{0i}F_{0i}\,\overline{X}X$. In the following we will denote the DM polarizability by $O_\gamma$, but the reader should keep in mind that when discussing specific UV completions with no large gap between the DM and the charged mediators masses another operator might be present. In addition, there may be $C$ and $P$ violating operators involving the Levi-Civita tensor, such as $F_{\mu\nu}\widetilde{F}^{\mu\nu}\,\overline{X}X$ and $F_{\mu\alpha}\widetilde{F}_{\nu\alpha}v^\mu v^\nu\,\overline{X}X$.

Self-conjugate DM also couples to photons via the anapole operator $\overline{X} s_\mu X\partial_\nu F^{\mu\nu}$, with $s_\mu$ the DM spin. This has lower dimensionality and generically dominates over $O_\gamma$ unless additional assumptions are made. Since the anapole violates separately $C$ and $P$, while $O_\gamma$ does not, a natural way to suppress its effects is to assume that the dark sector approximately respects either $C$ or $P$. Similarly, $\widetilde O_\gamma=\epsilon^{\mu\nu\alpha\beta}F_{\mu\nu} F_{\alpha\beta}i\overline{X}\gamma^5 X$ can dominate over the anapole if the dark sector is approximately $C/P$ invariant. We thus conclude that, under reasonable and generic conditions, the couplings  of self-conjugate DM to photons are controlled by $O_\gamma, \widetilde O_\gamma$.

Direct detection via DM polarizability was first studied in~\cite{Pospelov:2000bq} in the limit in which the interaction is described by a DM wave propagating in the electro-magnetic field of an infinitely heavy target nucleus. Later, the authors of ref.~\cite{Weiner:2012cb} emphasized that DM scattering for arbitrary masses proceeds via a photon loop, and estimated the rate using an effective field theory for the nucleus. More recent work on the DD signatures of (\ref{Ogamma}) can be found in~\cite{Frandsen:2012db} and~\cite{Crivellin:2014gpa}.

In this paper we present a detailed analysis of the DD signature induced by $O_\gamma$ and $\widetilde O_\gamma$. After a qualitative discussion of the nucleon/target effective field theory (EFT) in section~\ref{sec:EFT}, our main results for $O_\gamma$ are presented in sec.~\ref{sec:nucleonEFT}. A numerical study in section~\ref{sec:signatures} emphasizes the unique nature of the corresponding DD signature. A comparison between our results and the existing literature is given in Appendix~\ref{sec:vanish}. The operator $\widetilde O_\gamma$ is discussed in section~\ref{sec:Ogammatilde}. In section~\ref{sec:UV} we emphasize some important features that characterize UV complete models with unsuppressed $O_\gamma, \widetilde O_\gamma$, and comment on the coupling $\overline{X}XH^\dagger H$. A summary of our results is presented in section~\ref{conclusions}.

\section{EFT at a qualitative level}
\label{sec:EFT}

We start with an analysis of $O_\gamma$, whereas $\widetilde O_\gamma$ will be discussed later on.

There are two types of direct detection signatures that (\ref{Ogamma}) can lead to: an elastic scattering $XT\to XT$ (here $T$ stands for the target nuclei) or an inelastic process $XT\to XT\gamma$~\cite{Weiner:2012cb}. The first is numerically a loop effect. The latter process arises at tree-level, but its rate is suppressed at least by a factor $v^24\pi/\alpha\sim10^{-3}$ ($v$ is the incoming DM velocity) compared to the former, and is therefore completely negligible.

\subsection{The nucleon Lagrangian}
\label{sec:nucleon}

The rate for the elastic scattering $XT\to XT$ may be found exploiting the hierarchy of scales
\ba
q\lesssim Q_0\ll m_N\ll m_T,
\ea
with $q$ the momentum transfer, $1/Q_0$ the radius of $T$, $m_N$ the nucleon mass, and $m_T$ the target mass. 

One first performs the RG evolution from the new physics scale $\sim\Lambda$ to the scale $\sim m_c$. Here one finds that $O_\gamma$ mixes with the quark mass operators $O_q=\overline{q}HqX^2$ at one loop, and the latter with the gluon operator $O_G=\frac{\alpha_s}{4\pi}G_{\mu\nu}^2X^2$ via an additional QCD loop. In addition, one should take care of the top and bottom quark thresholds. Once this is done, the EFT at leading order in $q/m_c$ reads $\sum_{i=\gamma,u,d,s,G}\frac{C_i}{\Lambda^3}O_i$ where, up to $O(1)$ numbers, 
\ba
C_{q,G}(m_c)\sim C_{q,G}(\Lambda)+\frac{\alpha}{\pi}C_\gamma(\Lambda),
\ea
with $\alpha=e^2/4\pi$ the fine structure constant. In section~\ref{sec:UV} we argue that the natural expectation in realistic models is $C_{q,G}(\Lambda)\gtrsim\frac{\alpha}{\pi}C_\gamma(\Lambda)$, with $C_{q,G}(\Lambda)\sim\frac{\alpha}{\pi}C_\gamma(\Lambda)$ achievable under reasonable conditions.

The Wilson coefficients $C_{q,G,\gamma}(m_c)$ can be calculated using standard perturbation theory (see~\cite{Frandsen:2012db}\cite{Crivellin:2014gpa} for a discussion of the case $C_{q,G}(\Lambda)=0$). Alternatively, one can derive the leading non-derivative interactions of $X$ by simply observing that (\ref{Ogamma}) renormalizes the QED gauge coupling. By a formal redefinition $(A,e)\to(A_{\rm eff},e_{\rm eff})$, with
\ba\label{trick}
e_{\rm eff}^2(X)=e^2\left(1+4C_\gamma\frac{\overline{X}X}{\Lambda^3}+O(X^4/\Lambda^6)\right),
\ea
we can remove $X$ from the Lagrangian (up to momentum-suppressed terms). The EFT at the lower scale is now a function of $e_{\rm eff}(X)$, whereas by gauge invariance $eA=e_{\rm eff}A_{\rm eff}$ does not depend on the DM. This trick for example implies
\ba\label{trick1}
{\delta{\cal L}}_{m_t}\supset-\frac{\partial\log m_t}{\partial\log\alpha} m_t\overline{t}t~4C_\gamma(\Lambda)\frac{\overline{X}X}{\Lambda^3},
\ea
in agreement with an explicit loop analysis.

Next one should match the quark EFT onto a theory for the nucleons $N=n,p$. The leading DM couplings now are:
\ba\label{mN}
\delta{\cal L}_{m_N}&=&\sum_{i=\gamma,p,n}\frac{C_i}{\Lambda^3}O_i+O(q/m_N),
\ea
with $O_\gamma$ defined in (\ref{Ogamma}), and $O_N=m_N\overline{N}N\overline{X}X$ ($N=p,n$). It is understood that all couplings and operators are renormalized at $\sim m_N$. The remainder $O(q/m_N)$ also includes the chiral corrections discussed in~\cite{Prezeau:2003sv}\cite{Cirigliano:2012pq}.

Importantly, the coefficients $C_{p,n}(m_N)$ receive, besides the familiar contributions from $O_{u,d,s,G}(m_c)$ (see~\cite{Hill:2014yxa} for a recent NLO analysis), also a correction induced by $O_\gamma(m_c)$ of order:
\ba
\delta C_{p,n}(m_N)\sim\frac{\alpha}{\pi}C_\gamma(m_c). 
\ea
This latter RG effect can be seen, for example, proceeding along the lines discussed around (\ref{trick}). (We emphasize that both proton and the neutron masses are corrected by QED at 1-loop, so that $C_n(m_N)$ is also affected despite the neutron has no net charge.) The crucial difference compared to the RG evolution at higher scales is that now the analog of eq.(\ref{trick1}) is violated by non-negligible higher derivative operators of order $m_N^2/\Lambda_{\rm QCD}^2\sim1$. In terms of a heavy baryon EFT these higher-derivative operators correspond to $O(\alpha)$ corrections to the nucleon masses, and more generally the two-nucleon Lagrangian: their main effect is a modification of the pion-nucleon coupling at the percent level. Unfortunately, with our current knowledge of QCD we cannot determine the Wilson coefficients of these operators, and thus $C_{p,n}(m_N)$, with an accuracy better than $O(1)$, even under the assumption (unlikely, according to section~\ref{sec:UV}) that $C_{q,G}(\Lambda)=0$.

\subsection{EFT for the target nucleus}
\label{sec:nuclearEFT}

To determine the scattering rate for the process $XT\to XT$ one can proceed in two equivalent ways. The first is based on an EFT for the target nucleus defined at scales $\sim Q_0$, and will be qualitatively discussed in this subsection. The second, which is the one we will adopt in this paper, will be analyzed in section~\ref{sec:nucleonEFT}.

At scales $\mu\lesssim Q_0\ll m_N$ the target nucleus $T$ is effectively a point-like particle of mass $m_T\gg Q_0$ and one should be allowed to use a heavy nucleus Lagrangian. Up to $O(q/Q_0)$, the EFT at $\mu\sim Q_0$ includes $O_\gamma$ as well as the contact operator
\ba\label{RT}
\frac{\alpha}{4\pi}X^2\overline{T}T\left[Z^2Q_0+Zm_p+(A-Z)m_n\right],
\ea
where we ignored numerical coefficients for simplicity. The contact operator mixes with $O_\gamma$ at one-loop under the RG, as seen from arguments completely analogous to those discussed above. The terms of order $Z,A$ also receive corrections from $C_N(m_N)$ in $\delta{\cal L}_{m_N}$. 

From (\ref{RT}) one immediately reads a {{short distance}} contribution to the amplitude for $XT\to XT$. There is also a correction coming from a UV-sensitive one-loop diagram involving $O_\gamma$, which scales as the $O(Z^2)$ term in (\ref{RT})~\cite{Weiner:2012cb}. The two contributions are individually scheme-dependent; only their sum is physical. For example, using a mass-independent renormalization scheme the loop diagram vanishes at $q=0$, and the $O(Z^2)$ effect comes dominantly from the counterterm (\ref{RT}).~\footnote{\label{foot}The authors of~\cite{Weiner:2012cb} neglected the contact operator (\ref{RT}), or in other words assume a certain renormalization scheme in which its coefficient vanishes. However, this is not necessarily the same scheme that the authors used to regulate the 1-loop diagram. This introduces a spurious scheme-dependence and an $O(1)$ uncertainty in the amplitude.}

\section{The 2-body process}
\label{sec:nucleonEFT}

The approach followed in section~\ref{sec:nuclearEFT} is intuitive from a physical standpoint, but not very convenient. One reason is that it depends on several unknown form-factors, even in the optimistic (and unrealistic) case in which only $O_\gamma$ is present at $\mu\sim m_N$. More importantly, though, it obscures the accuracy of the perturbative expansion. For example, are we allowed to ignore QED vertex corrections to the one-loop diagram of~\cite{Weiner:2012cb}? These are naively of order $\alpha Z^2/4\pi$, and apparently not negligible for heavy targets.

In this section we will approach the problem from the point of view of the ``fundamental" nucleon EFT. In practice we take (\ref{mN}) as our starting point, derive a multi-body effective theory for the nucleons, and finally take the appropriate nuclear matrix element. Using this formalism all the unknowns will be encoded in measurable nuclear form factors. Furthermore, within this formalism the perturbative expansion becomes manifest. For instance, an inspection of the $O(\alpha Z^2/4\pi)$ ``vertex corrections" mentioned above shows that these are secretly a renormalization of the nuclear wave-function, and hence already included in the nuclear potential.

\subsection{The 2-proton form factor}
\label{sec:2body}

We now want to calculate the amplitude for $XT\to XT$ from the {\emph{nucleon}} Lagrangian (\ref{mN}). The discussion in section~\ref{sec:nuclearEFT} shows that this process receives contributions from the operators $O_{n,p}$ in (\ref{mN}) as well as $O_\gamma(m_N)$. The former may be treated using standard methods. Our main focus here will be on $O_\gamma$. In section~\ref{sec:signatures} we will study in detail the interplay between all Wilson coefficients $C_{\gamma,p,n}$.

To proceed we formally write a multi-nucleon hamiltonian $H_{\rm tot}=H_{\rm strong}+V$, where $H_{\rm strong}$ contains the nuclear force and $V$ one insertion of $O_\gamma$. At leading order in the weak DM coupling and all orders in the nuclear force the amplitude for $XT \to XT$ is just the Born approximation 
$$
\langle T_f|V|T_i\rangle, 
$$
with the nuclear ground states $|T_{i,f}\rangle$ understood as $A$-nucleon configurations dressed with the nuclear force.

\begin{figure}[t]
\begin{center}
\includegraphics[width=6cm]{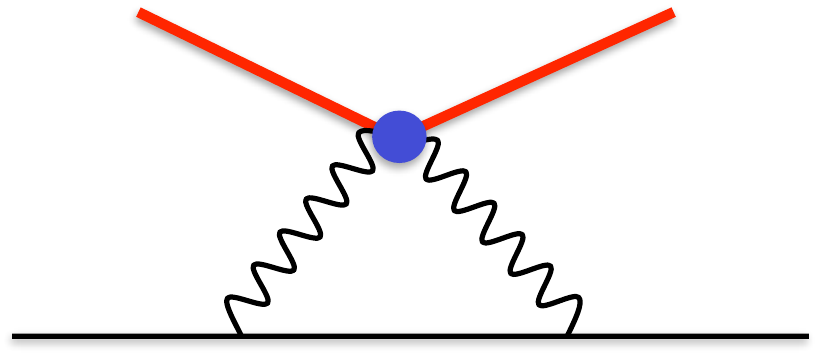}~~~~~~~~\includegraphics[width=6cm]{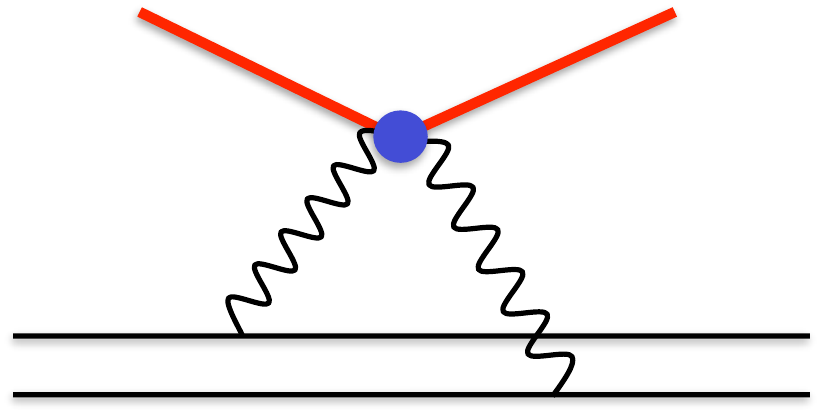}
\caption{Feynman diagrams for the 1-proton and 2-proton processes. Both diagrams contribute to the contact DM-nuclei interaction, whereas the one on the right is also related in a scheme-dependent way to the one-loop diagram of~\cite{Weiner:2012cb} (see Appendix~\ref{sec:vanish} for details).}\label{fig:FD}
\end{center}
\end{figure}

The dominant DM-nucleon interactions contributing to the potential $V$ are described by the diagrams shown in fig.~\ref{fig:FD}. The loop on the left vanishes for $q=0$ in any mass-independent renormalization scheme, which is the natural regulator in our EFT. Therefore, only the 2-body process in the right of figure~\ref{fig:FD} is relevant at leading $q/m_N$ order. 

The non-relativistic amplitude for the 2-proton process $ppX\to ppX$ is  
\ba\label{1}
\overline{\cal M}_2({\bf q}_i,{\bf q}_j)=\delta_{s_is_i'}\delta_{s_js_j'}V_0\frac{{\bf q}_i\cdot {\bf q}_j}{{\bf q}_i^2{\bf q}_j^2}(1+O({\bf q}^2/m_N^2)), ~~~~~~~~V_0=-8\frac{e^2}{\Lambda^3}C_\gamma(m_N),
\ea
where ${\bf q}_{i,j}$ are the three-momentum transferred to the nucleons, whereas $s_{i,j,i',j'}$ are spin indices for the nucleons (the DM spin indices are not shown because they cancel out in the cross section when summing and averaging over final and initial states). We did not add the crossed diagram because it will be automatically included when convoluting $\overline{\cal M}_2$ with the anti-symmetric nuclear wave function. Following~\cite{Prezeau:2003sv}\cite{Cirigliano:2012pq}, we used a non-relativistic normalization for the 1-particle states. In practice, this corresponds to divide the relativistic amplitude by $(2m_N)^2(2m_X)$. With this convention, the formula (\ref{1}) also applies to real scalars (with $C_\gamma$ a parameter with dimensions of a mass).

The DM-nucleon potential in ``mixed coordinates" ($x_i$ for nucleon $i$ and Fourier for $X$) reads:
\ba
\tilde V_{ij}&=&-\int\frac{d{\bf q}_i}{(2\pi)^3}\int\frac{d{\bf q}_j}{(2\pi)^3}e^{-i{\bf q}_i\cdot {\bf x}_i-i{\bf q}_j\cdot {\bf x}_j}(2\pi)^3\delta^{(3)}({\bf q}+{\bf q}_i+{\bf q}_j)\overline{\cal M}_2({\bf q}_i,{\bf q}_j)\\\no
&=&\delta_{s_is_i'}\delta_{s_js_j'}V_0e^{+i{\bf q}\cdot {\bf R}}f({\bf q},{\bf r}),
\ea 
where ${\bf R}=({\bf x}_i+{\bf x}_j)/2$, ${\bf r=x}_i-{\bf x}_j$, and
\ba\label{f}
f({\bf q},{\bf r})&=&\int\frac{d{\bf k}}{(2\pi)^3}e^{-i{\bf k}\cdot {\bf r}}\frac{\left({\bf k}-\frac{\bf q}{2}\right)\cdot \left({\bf k}+\frac{\bf q}{2}\right)}{\left({\bf k}-\frac{\bf q}{2}\right)^2\left({\bf k}+\frac{\bf q}{2}\right)^2}\\\no
&=&\frac{1}{4\pi r}\int_{-1/2}^{+1/2}dy~e^{-qr\sqrt{\frac{1}{4}-y^2}}\left[\left(1-qr\sqrt{\frac{1}{4}-y^2}\right)\cos(y{\bf q}\cdot{\bf r})-(y{\bf q\cdot r})\sin(y{\bf q\cdot r})\right]\\\no
&=&\frac{1}{4\pi r}\left[1-\frac{\pi}{4}qr+\frac{1}{4}(qr)^2-\frac{1}{8}({\bf q}\cdot{\bf r})^2+O(q^3r^3)\right].
\ea
The term linear in $q$ arises because the transition is mediated by a massless particle.

Because $V=\sum_{i<j}\tilde V_{ij}$, we conclude that the amplitude for $XT\to XT$, with the target remaining in the ground state, is given by~\cite{Prezeau:2003sv}\cite{Cirigliano:2012pq}
\ba\label{TV}
\langle T_f|\sum_{i<j}\tilde V_{ij}|T_i\rangle&=&\sum_{i<j}\int d{\bf x}_i\int d{\bf x}_j~\tilde V_{ij}\otimes{\hat\rho}^{(2)}({\bf x}_i,{\bf x}_j)\\\no
&=&\frac{Z(Z-1)}{2}\int d{\bf x}_1\int d{\bf x}_2~\tilde V_{12}\otimes{\hat\rho}^{(2)}({\bf x}_1,{\bf x}_2).
\ea
Here ${\hat\rho}^{(2)}({\bf x}_i,{\bf x}_j)$ is the (diagonal) 2-proton nuclear density matrix, the sum extends over all proton pairs, and $\otimes$ indicates a contraction of the spin indices. In the second line we used the fact that protons are indistinguishable. The factor $Z(Z-1)$ signifies that this is truly a two-body effect, and as such it vanishes for $Z=1$. The relevant spin-singlet quantity is the projection:
 \ba
\rho^{(2)}({\bf x}_i,{\bf x}_j)&\equiv& (\delta_{s_is_i'}\delta_{s_js_j'})\otimes\hat\rho^{(2)}({\bf x}_i,{\bf x}_j)\\\no
&=&\int \frac{d{\bf q}_1}{(2\pi)^3} \int \frac{d{\bf q}_2}{(2\pi)^3} e^{-i{\bf q}_1\cdot {\bf x}_1-i{\bf q}_2\cdot {\bf x}_2}F^{(2)}({\bf q}_1,{\bf q}_2).
 \ea

Finally, employing eq. (\ref{f}), and making a trivial coordinate transformation, we arrive at an expression for the dominant, spin-independent part of the amplitude:
\ba\label{M2b}\label{MNR}
\langle T_f|\sum_{i<j}\tilde V_{ij}|T_i\rangle&=&\frac{Z(Z-1)}{2}V_0~F_{pp}(q),
\ea
where we defined the {\emph{2-proton form factor}}~\footnote{If we were to follow the conventions used in~\cite{Prezeau:2003sv}\cite{Cirigliano:2012pq} we would call this quantity $F_{\gamma\gamma}$, because induced by the exchange of two photons.}
\ba\label{2body}
F_{pp}(q)=\int\frac{d{\bf k}}{(2\pi)^3}\frac{\left({\bf k}-\frac{\bf q}{2}\right)\cdot \left({\bf k}+\frac{\bf q}{2}\right)}{\left({\bf k}-\frac{\bf q}{2}\right)^2\left({\bf k}+\frac{\bf q}{2}\right)^2}F^{(2)}(-{\bf k}+{\bf q}/2,{\bf k}+{\bf q}/2).
\ea
Eqs. (\ref{MNR}) and (\ref{2body}) are one of the main result of the present paper.

Note that the relative size between two-body and one-body contributions induced by $O_{p,n}$ is $\sim\frac{ZQ_0}{m_N}$ (see also eq.(\ref{RT})), which is not negligible for heavy nuclei. On the other hand, consistently with what done in (\ref{mN}) we can neglect corrections $O(q^2/m_N^2)$ to both the one-body and two-body terms because of order $Q_0^2/m_N^2=$ few \%.~\footnote{This means that $Z(Z-1)$ should truly be replaced by $Z^2$ in eq.(\ref{MNR}). We will do this in section~\ref{sec:signatures} when we will include the corrections from $C_N(m_N)$. For now we decided to keep $Z(Z-1)$ in (\ref{MNR}) to emphasize the 2-body nature of the amplitude.} 

\subsection{The role of proton-proton correlations}
\label{sec:corre}

The 2-proton density $F^{(2)}$ appearing in (\ref{2body}) plays an important role in many nuclear reactions, such as electro-disintegration processes, $(e,e'N)$ and $(e,e'NN)$, precision calculations in muonic atoms, and neutrino-nucleus interactions.

In~\cite{Prezeau:2003sv}\cite{Cirigliano:2012pq}\cite{Cirigliano:2013zta}, the relevance of 2-body densities in DM detection has been pointed out, although as a subleading effect in the chiral counting. In this paper we find another interesting application for DM direct detection, where the two-body term plays a dominant role.

The exact form of $F^{(2)}$ is not known, but some of its basic properties can be qualitatively understood. Without loss of generality we write
\ba\label{corr}
F^{(2)}(-{\bf k}+{\bf q}/2,{\bf k}+{\bf q}/2)=F_p^{(1)}(-{\bf k}+{\bf q}/2)F_p^{(1)}({\bf k}+{\bf q}/2)+F_{\rm corr}({\bf k},{\bf q}),
\ea
where $F_p^{(1)}$ is the (one-proton) charge form factor, and $F_{\rm corr}$ a measure of the correlation between the two nucleons (protons in our case). From the normalization of the density distributions follows that $F_{\rm corr}(0,0)=0$. In practice this means that one may neglect the correlation when {\emph{both}} $|{\bf q}|,|{\bf k}|$ are much smaller than, say, the pion mass $m_\pi$. As ${\bf q}^2,{\bf k}^2\sim m^2_\pi$, nucleon-nucleon correlations become non-negligible and the approximation $F_{\rm corr}=0$ is violated. At even larger momenta one anticipates a universal shape for $F^{(2)}$, dominated by the repulsive pion exchange.

From this simple consideration follows that $F_{\rm corr}$ cannot be ignored in general, because the integral in (\ref{2body}) probes a regime where the correlation is presumably non-negligible. In particular, $F_{\rm corr}({\bf k},0)$ is likely to result in an overall $O(1)$ correction in the nuclear matrix element. To see this more explicitly, we take the following phenomenological expression for the charge form factor
\ba\label{pheno}
F^{\rm pheno}_p=e^{-\bar{q}^2},~~~~~~\bar q=|{\bf q}|/Q_0,
\ea
and parametrize $F^{(2)}$ with
$$
F^{(2), {\rm pheno}}=F^{\rm pheno}_p(-{\bf k}+{\bf q}/2)F^{\rm pheno}_p({\bf k}+{\bf q}/2)\left[1+c_1\frac{{\bf k}^2}{Q_0^2}+c_2\frac{{\bf q}^2}{Q_0^2}+c_3\frac{{\bf k}\cdot{\bf q}}{Q_0^2}\right], 
$$
where $c_i$ are order one numbers (in general functions of $A,Z$) mimicking the effect of a short distance correlation. The expression (\ref{2body}) can now be solve exactly:
\ba\label{phenoFpp}
F_{pp}^{\rm pheno}&=&\frac{Q_0}{4\sqrt{2}\pi^{3/2}}\overline F_{pp}^{\rm pheno},\\\no
\overline F_{pp}^{\rm pheno}&=&\frac{Q_0^2}{4}e^{-q^2/2Q_0^2}\int d{\bf r}~e^{-{\bf r}^2Q_0^2/8}f({\bf q},{\bf r})\left[\left(1+\frac{3}{4}c_1\right)-\frac{1}{16}c_1{\bf r}^2Q_0^2+c_2\frac{{\bf q}^2}{Q_0^2}\right]\\\no
&=&e^{-\bar q^2/2}\left[1+\frac{1}{4}c_1-\frac{\pi^{3/2}}{2\sqrt{2}}\bar q+\left(\frac{5}{3}-\frac{5}{12}c_1+c_2\right)\bar q^2+O(\bar q^3)\right].
\ea
In the second line we used the expansion of $f$ given in (\ref{f}). As expected, we see that $c_1$ changes the overall rate, whereas both $c_{1,2}$ modify the momentum-dependence of the form factor ($c_3$ does not contribute for our choice of $F^{(2), {\rm pheno}}$). We will present a numerical study in the next section. Interestingly, when $c_i\to0$ the form factor $\overline F_{pp}^{\rm pheno}$ reduces to that derived in~\cite{Weiner:2012cb}\cite{Frandsen:2012db}. This correspondence is elucidated in Appendix~\ref{sec:vanish}.

We conclude this section observing that the proton-proton correlation becomes parametrically small if one models the nuclear potential with a mean field approximation. Indeed, in that case the nuclear wave-function is described by a single Slater determinant, and it is a trivial exercise to show that $F_{\rm corr}=O(1/Z)$ is entirely due to the Pauli exclusion principle. (We also numerically verified this expectation using a shell model.) In reality the mean field potential is no more than an intuitive picture of the nucleus, and $F_{\rm corr}$ cannot be neglected.

\section{Signatures in direct detection experiments}
\label{sec:signatures}

The study of the DD signature induced by $O_\gamma$ is complicated by two obvious hurdles. First, we do not have a reliable estimate of the 2-proton form factor (\ref{2body}). Second, large hadronic uncertainties make it impossible to precisely determine the actual relation among the Wilson coefficients in (\ref{mN}) and the fundamental parameters. 

We thus take a bottom-up, phenomenological approach. For the form factors we use the phenomenological expressions given in (\ref{pheno}) and (\ref{phenoFpp}). The spin-independent differential rate is then given by
\ba\label{SI}
\frac{d\sigma_T}{dE_R}=\frac{\mu_T^2}{\pi E_R^{\rm max}}\left(\frac{m_N}{\Lambda^3}C_p\right)^2\left\{\frac{f_{pp}}{f_p}\frac{Q_0}{m_N}Z^2\overline F_{pp}^{\rm pheno}(q)+\left[{Z}+\frac{f_n}{f_p}(A-Z)\right]F^{\rm pheno}_{p}(q)\right\}^2,
\ea
with
\ba
\frac{f_{pp}}{f_p}=2\sqrt{\frac{2}{\pi}}\alpha\frac{C_\gamma}{C_p}\frac{Z-1}{Z},~~~~~~~~~~~f_N=C_N\frac{m_N}{\Lambda^3}.
\ea
We simplified eq.(\ref{SI}) assuming that protons and neutrons have the same mass and 1-particle densities. We use $Q_0=0.5(0.3 + 0.9A^{1/3})^{-1}$ GeV.

Now, the discussion in section~\ref{sec:UV} (see also section~\ref{sec:nucleon}) suggests that, in generic theories with unsuppressed $C_\gamma$, the natural expectation is $C_N/C_\gamma\sim\alpha/\pi$, and hence $f_{pp}/f_p, f_n/f_p=O(1)$. In the following we will therefore treat these latter as independent parameters of order unity. Because $f_{pp}$ already accounts for changes in the overall normalization of the two-body effect, it makes sense to work with form factors that are normalized to one at $q=0$; for this reason we will set $c_1=0$ and vary only $c_2$ in our numerical analysis ($c_3$ has no effect on (\ref{phenoFpp})). Comparing to~\cite{Frandsen:2012db}, our phenomenological 2-body form factor for $c_1\to0$ reduces to $\overline F_{pp}^{\rm pheno}\to F_{\rm Ray}(1+c_2\bar q^2)$ (see Appendix~\ref{sec:vanish}).

\subsection{Numerical analysis}

The impact of the new form factor may be significant, both in the spectrum and the total rate.

When both $f_{pp}/f_p, f_{pp}/f_n$ have the same magnitude and sign, the new term $f_{pp}$ is expected to dominate for heavy nuclei (at least as long as the form factors are positive and non-vanishing). However, an opposite sign in either $f_{pp}/f_p, f_n/f_p$ can generically result in destructive interference among the various contributions to (\ref{SI}), and hence lead to qualitatively new effects.~\footnote{This possibility has already been noticed in~\cite{Frandsen:2012db} for a particular set of $f_{pp,p,n}$ and a vanishing 2-proton correlation ($c_i=0$).} Because of the different momentum-dependence of the form factors, the suppression in $d\sigma_T/dE_R$ will occur at a specific recoil energy. This energy is a strong function of the parameters $f_{pp}/f_p, f_n/f_p$ and the form factors, as well as of the target, due to the peculiar dependence on $A,Z$.

To assess the effect of $f_{pp}/f_p, f_n/f_p$ on the total rate, the relevant quantity to consider is the number of events within a certain signal region $\Delta$:
\ba\label{ND}
N_\Delta=\frac{\rho_X}{m_X}~{\rm Ex}\sum_{\rm Target}N_T\int_{\Delta}ds~\eta_{\rm eff}(s)\int dE_R~p(E_R,s)\int_{v_{\rm min}(E_R)} d{\bf v}~v{f_{\rm lab}({\bf v,v_e})}~\frac{d\sigma_T}{dE_R}.
\ea
Here $\rho_X$ is the local DM density, Ex the detector exposure, $N_T$ the number of nuclear targets (summed over all isotopes), $\eta_{\rm eff}(s)p(E_R,s)$ the signal efficiency, and $f_{\rm lab}$ the DM velocity distribution in the lab frame.

While for $f_{pp}=0$ the surface $N_\Delta={\rm const}$ is defined by parallel lines $f_n\propto f_p$, in the more general case it becomes an ellipses. To see this one can focus on a single isotope, in which case, denoting by $\langle\rangle$ the integrals in (\ref{ND}), it is clear that
\ba\label{CS}\no
{N_\Delta(f_{pp},f_p,f_n)}&=&\langle(x {\overline F_{pp}}+yF_{p})^2\rangle\\
&=&\langle F_{p}^2\rangle\left(y+x\frac{\langle {\overline F_{pp}}F_{p}\rangle}{\langle F_{p}^2\rangle}\right)^2+\langle\overline F_{pp}^2\rangle\left(1-\frac{\langle {\overline F_{pp}}F_{p}\rangle^2}{\langle \overline F_{pp}^2\rangle\langle F_{p}^2\rangle}\right)x^2,
\ea
for $x\propto f_{pp}/f_p$ and $y\propto (\frac{Z}{A-Z}+f_n/f_p)$. By the Cauchy-Schwarz inequality the last term is always positive, implying that the isocurves must be ellipses.

\begin{figure}[h]
\begin{center}
\includegraphics[width=7.5cm]{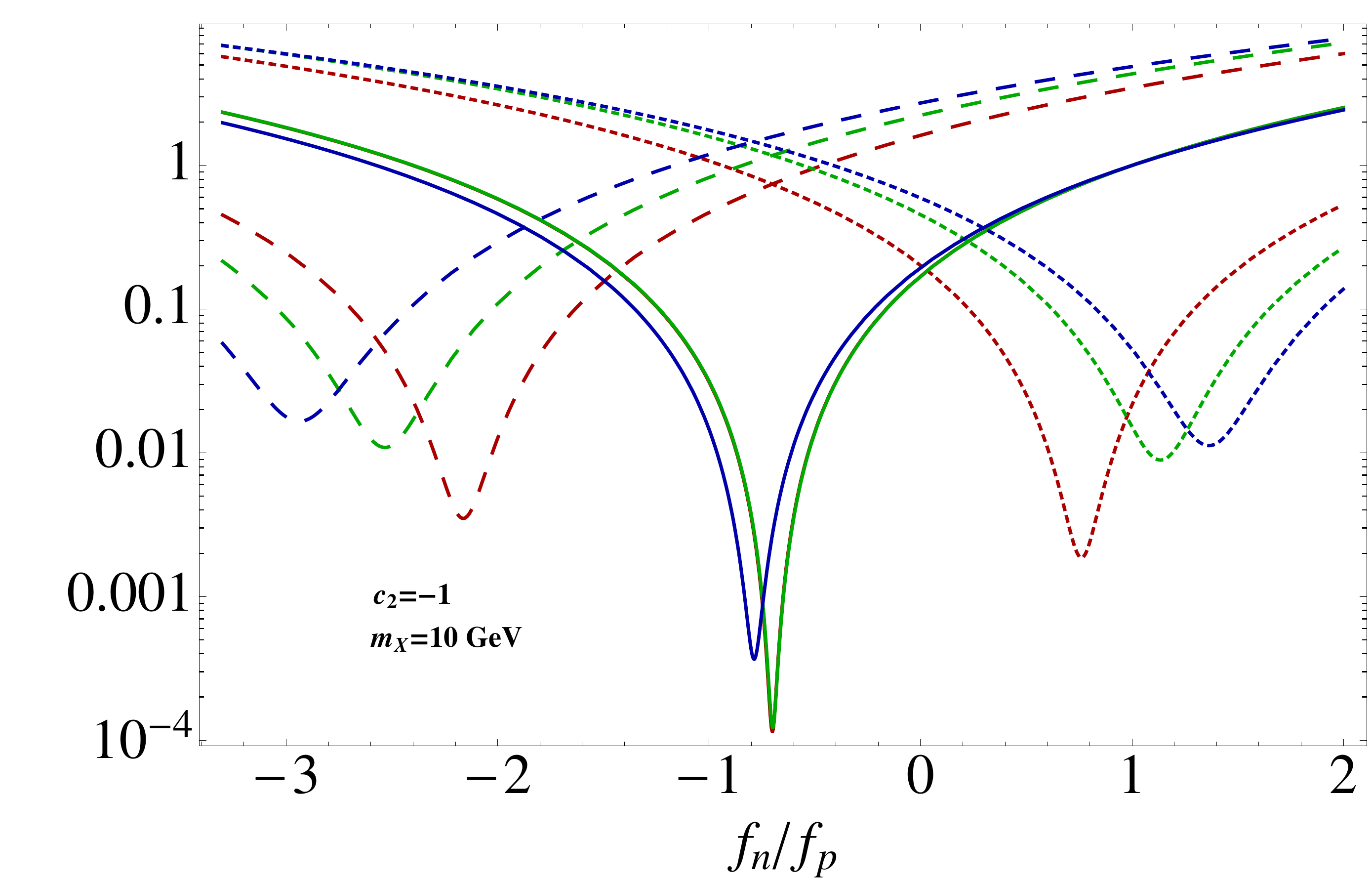}\includegraphics[width=7.5cm]{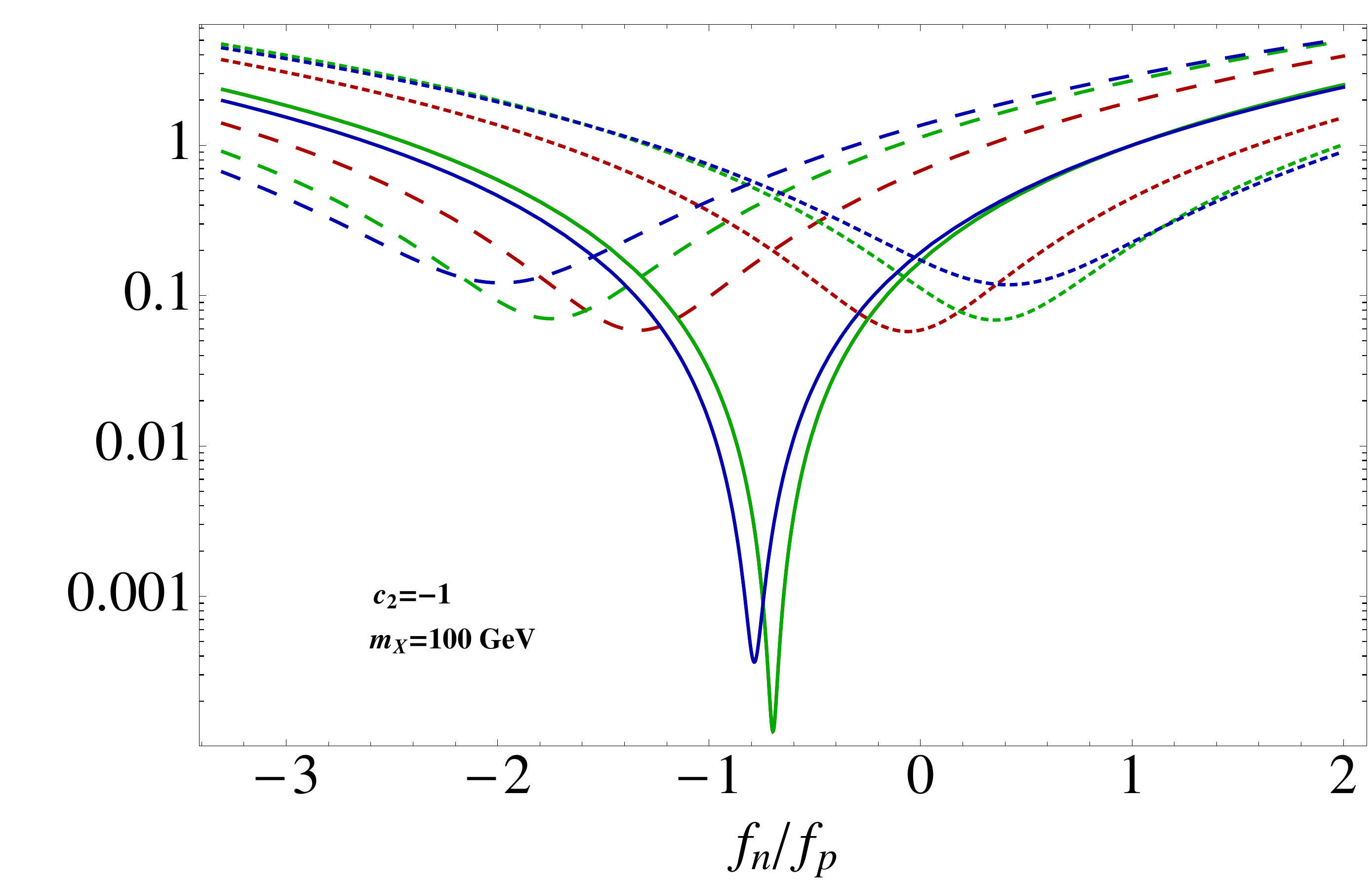}
\includegraphics[width=7.5cm]{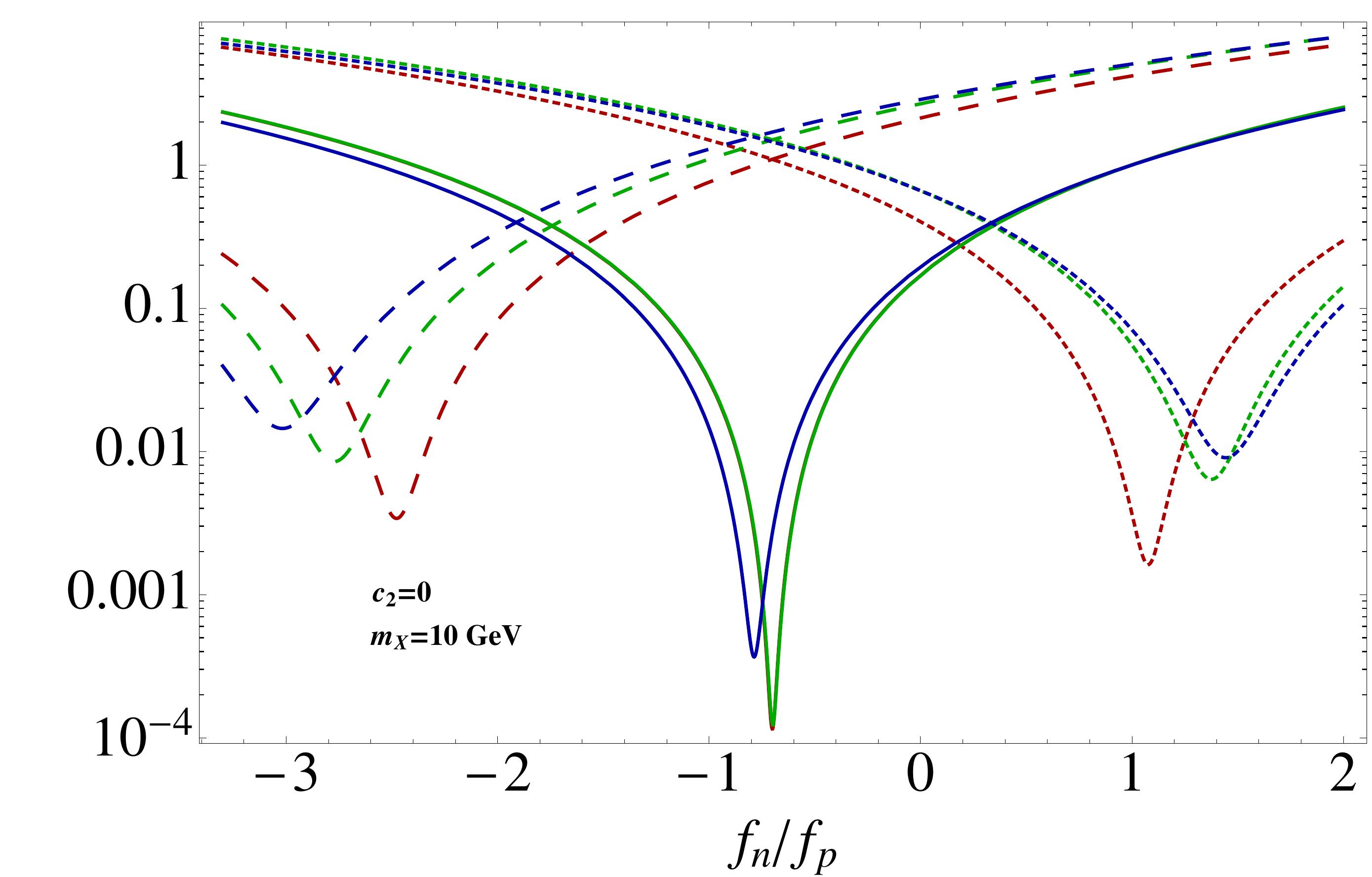}\includegraphics[width=7.5cm]{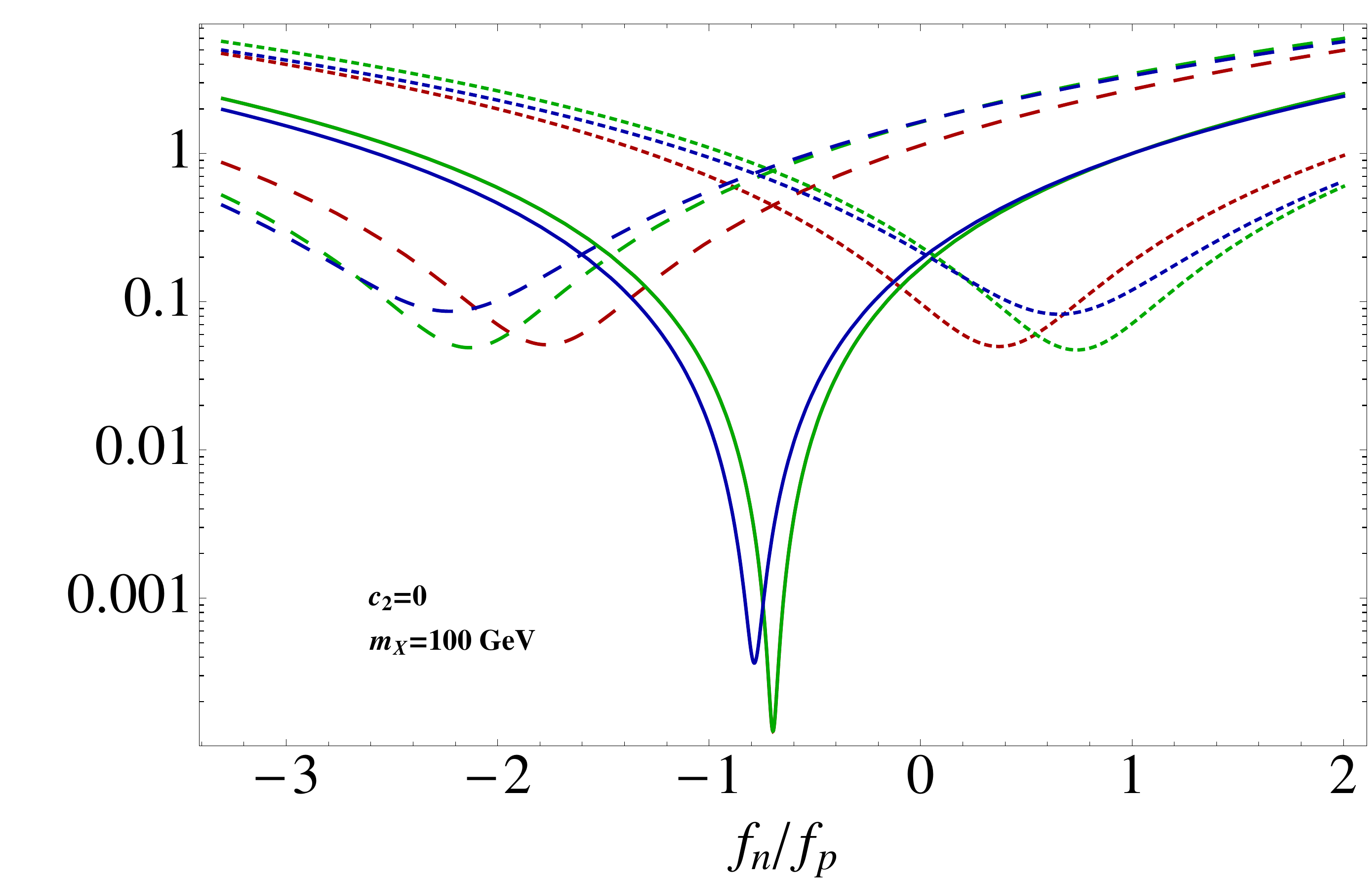}
\includegraphics[width=7.5cm]{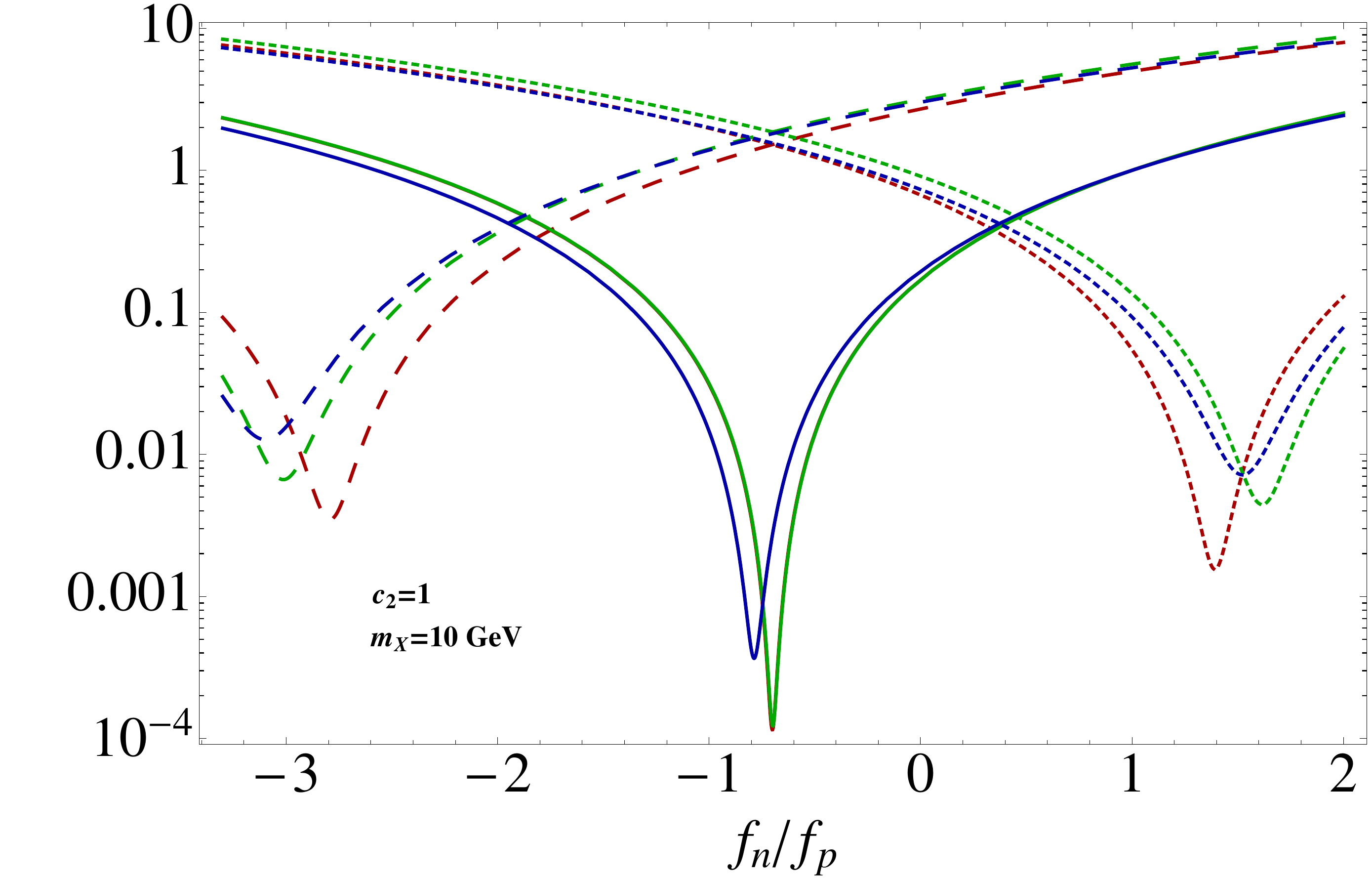}\includegraphics[width=7.5cm]{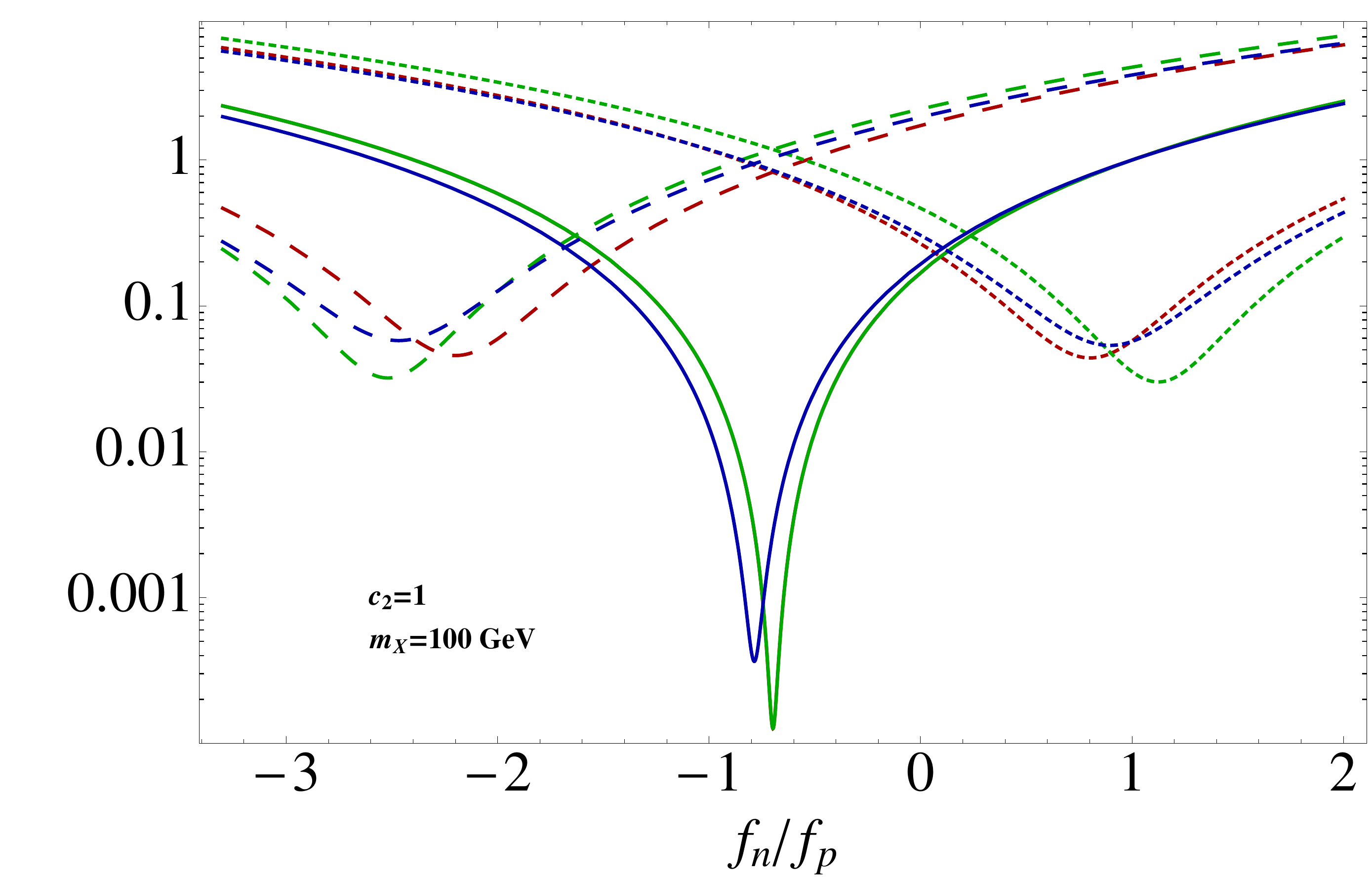}
\caption{Degradation plot ($N_\Delta/N^{\rm ref}_\Delta$) for $f_{pp}/f_{p}=-1$ (dotted lines) $f_{pp}/f_{p}=0$ (solid) and $f_{pp}/f_{p}=1$ (dashed). Left column corresponds to $m_X=10\text{ GeV}$, and right column to $m_X=100\text{ GeV}$. Color coding: Red (Xenon100), Green (LUX), Blue (CDMS-Ge). Note that for $f_{pp}=0$ the solid red and green lines exactly overlap. }\label{fig:interplay}
\end{center}
\end{figure}

\begin{figure}[h]
\begin{center}
\includegraphics[width=6.35cm]{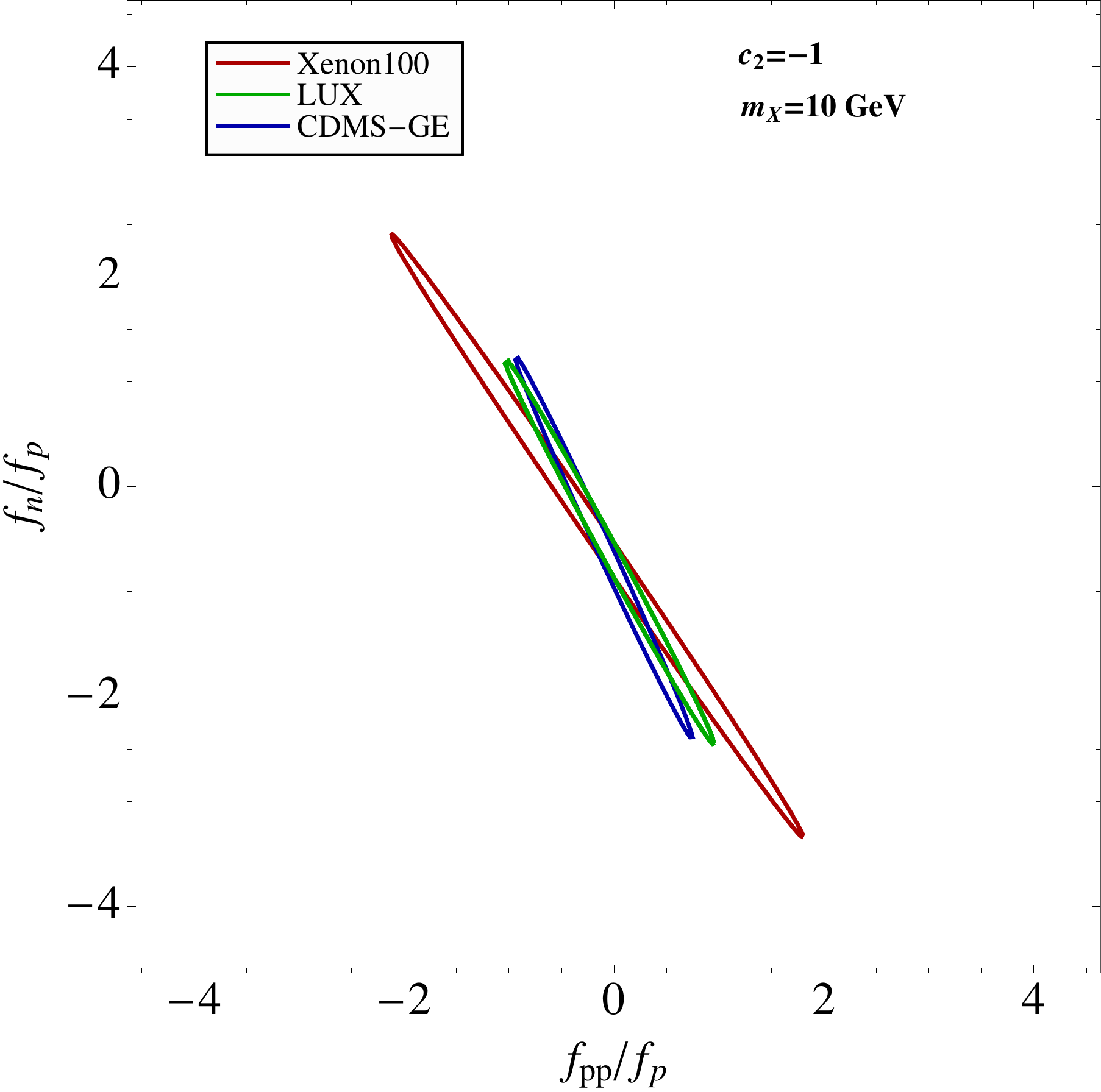}~~~~~~~\includegraphics[width=6.35cm]{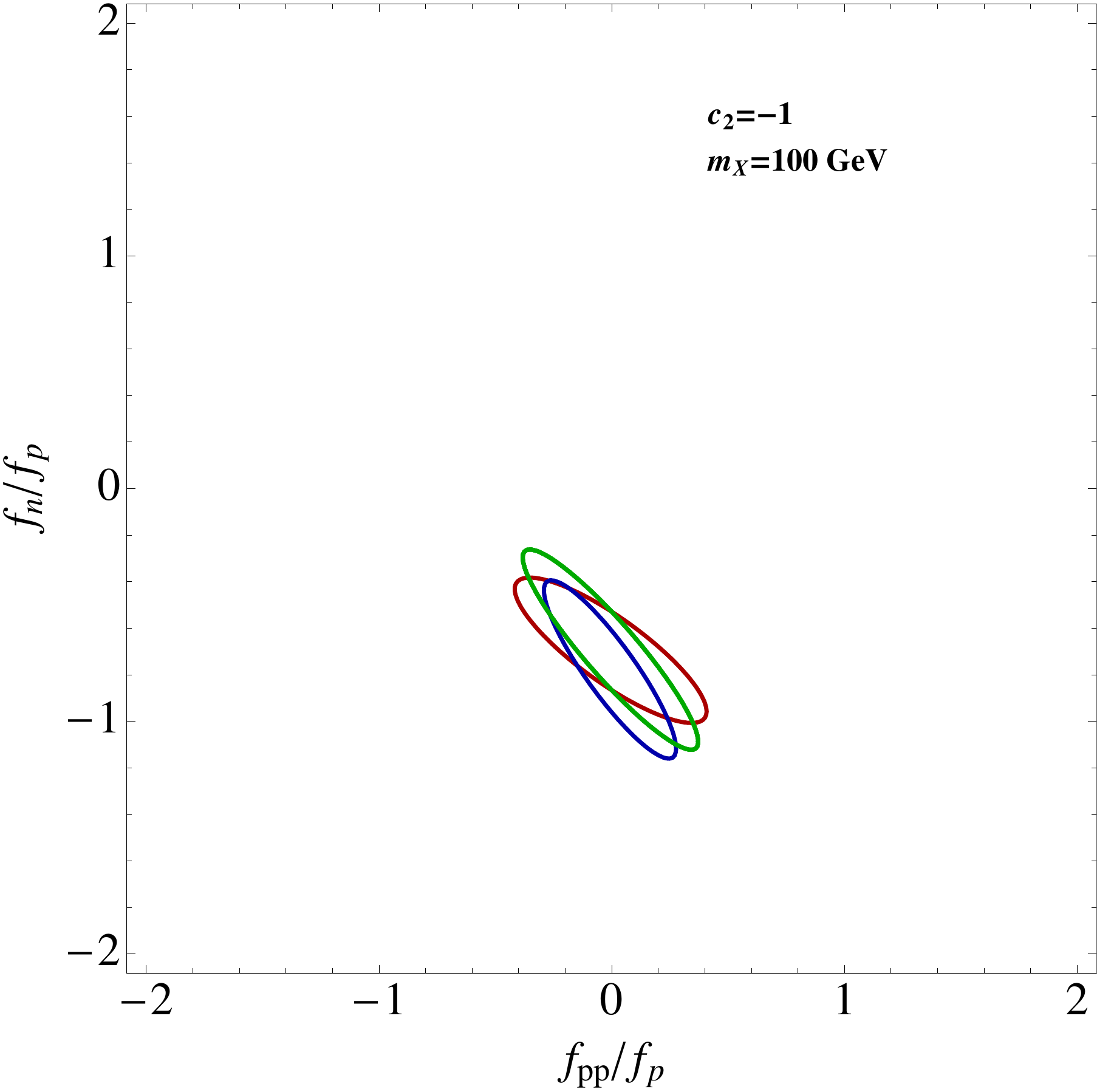}
\includegraphics[width=6.35cm]{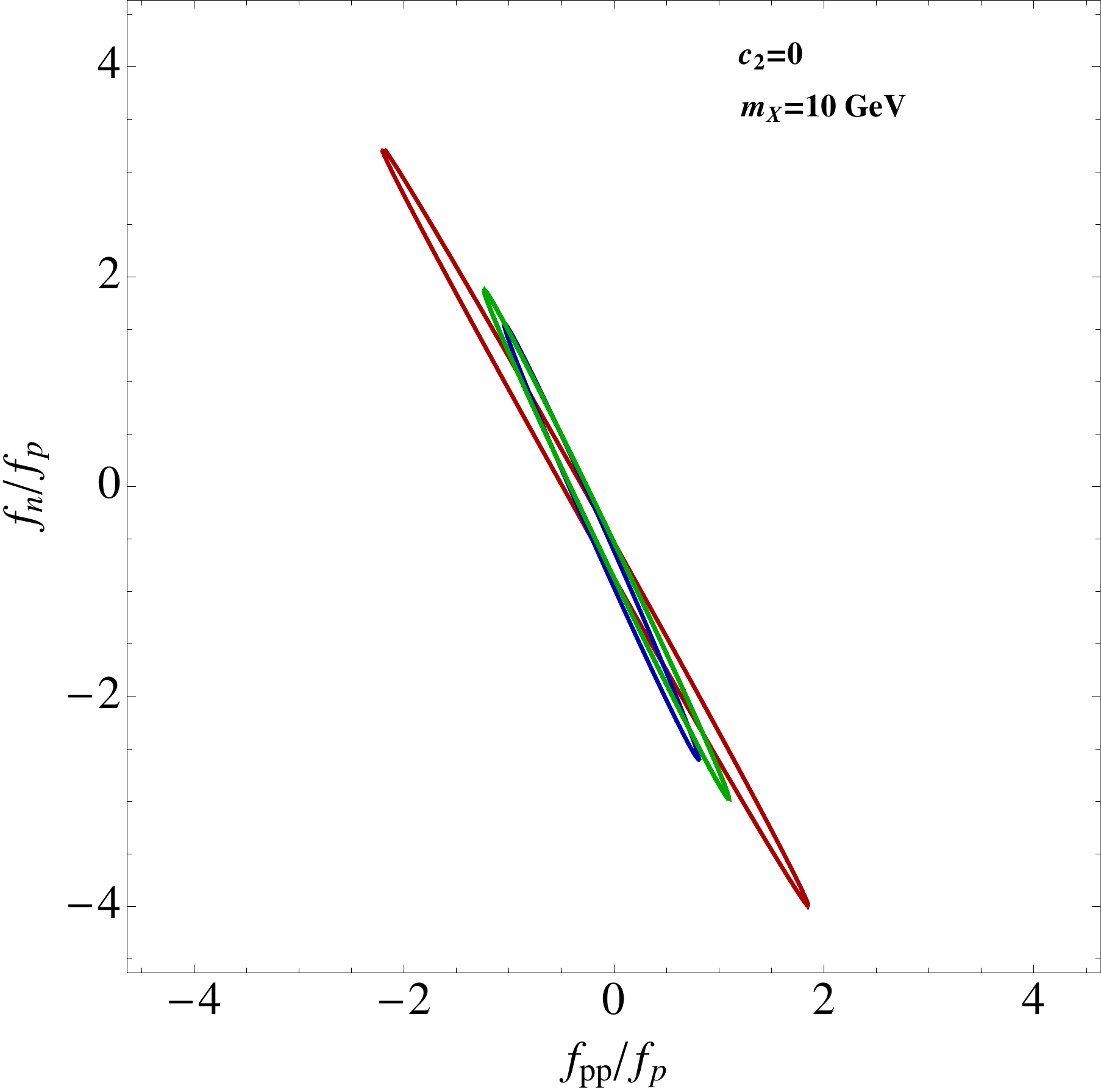}~~~~~~~\includegraphics[width=6.35cm]{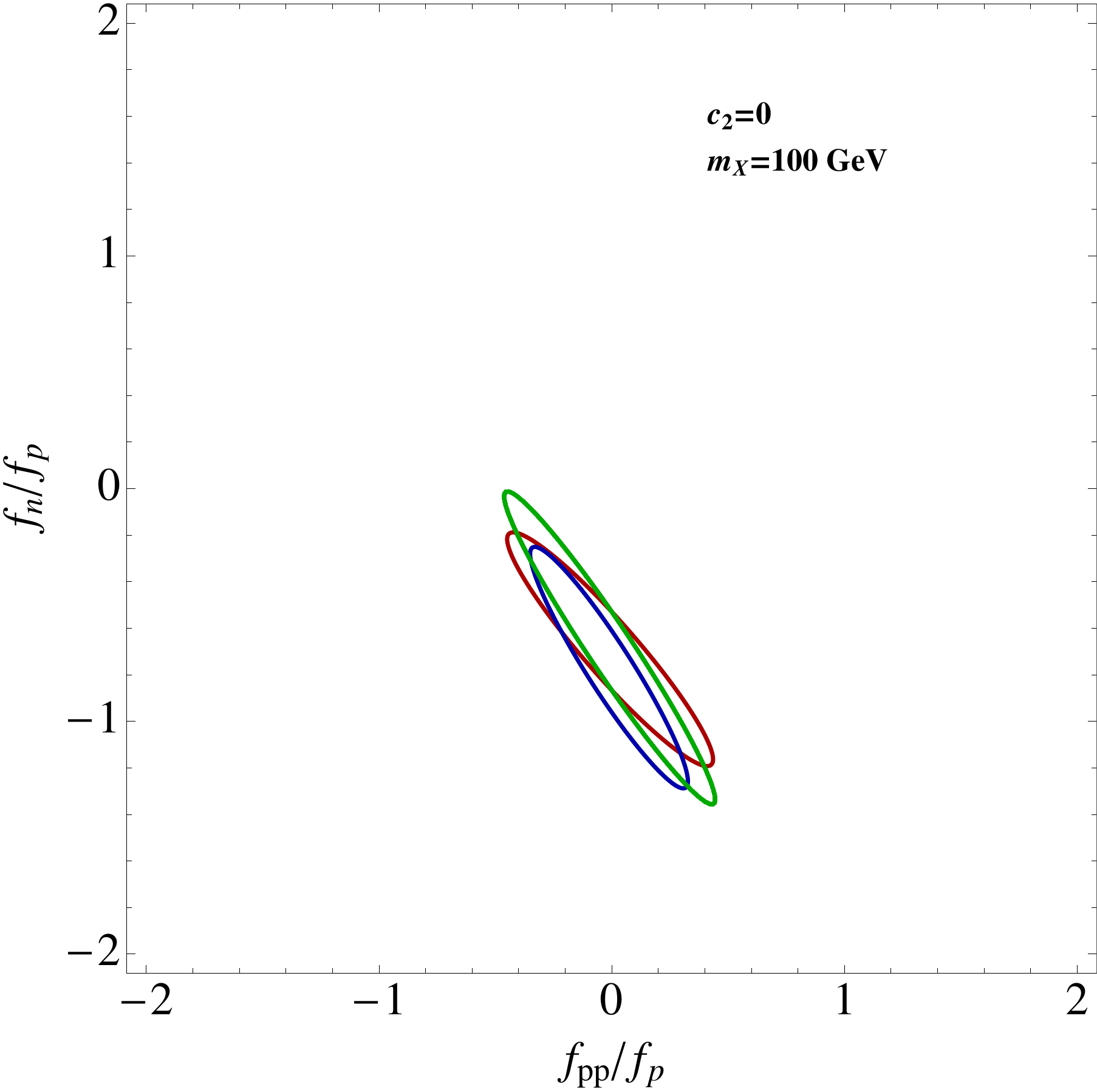}
\includegraphics[width=6.35cm]{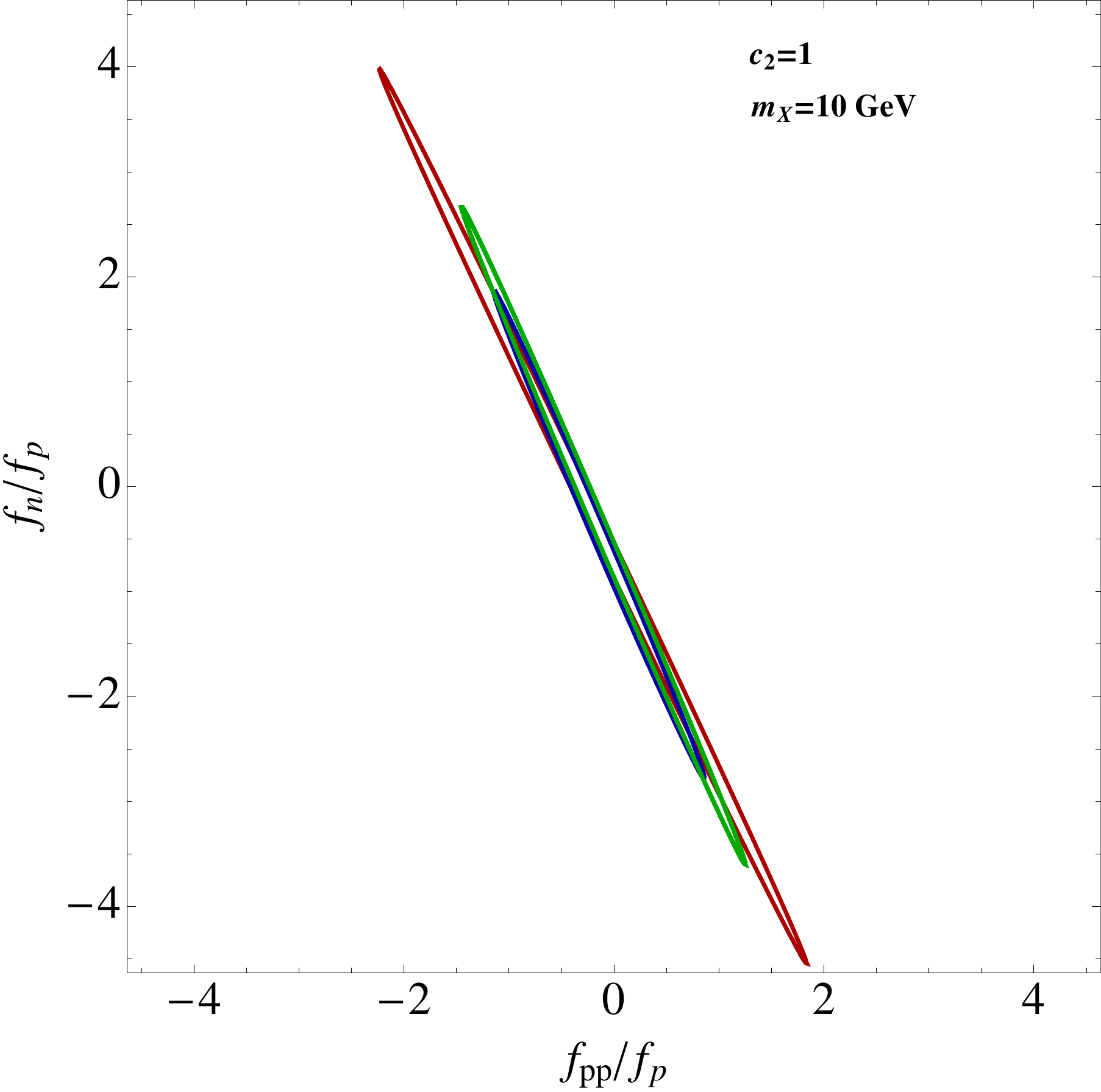}~~~~~~~\includegraphics[width=6.35cm]{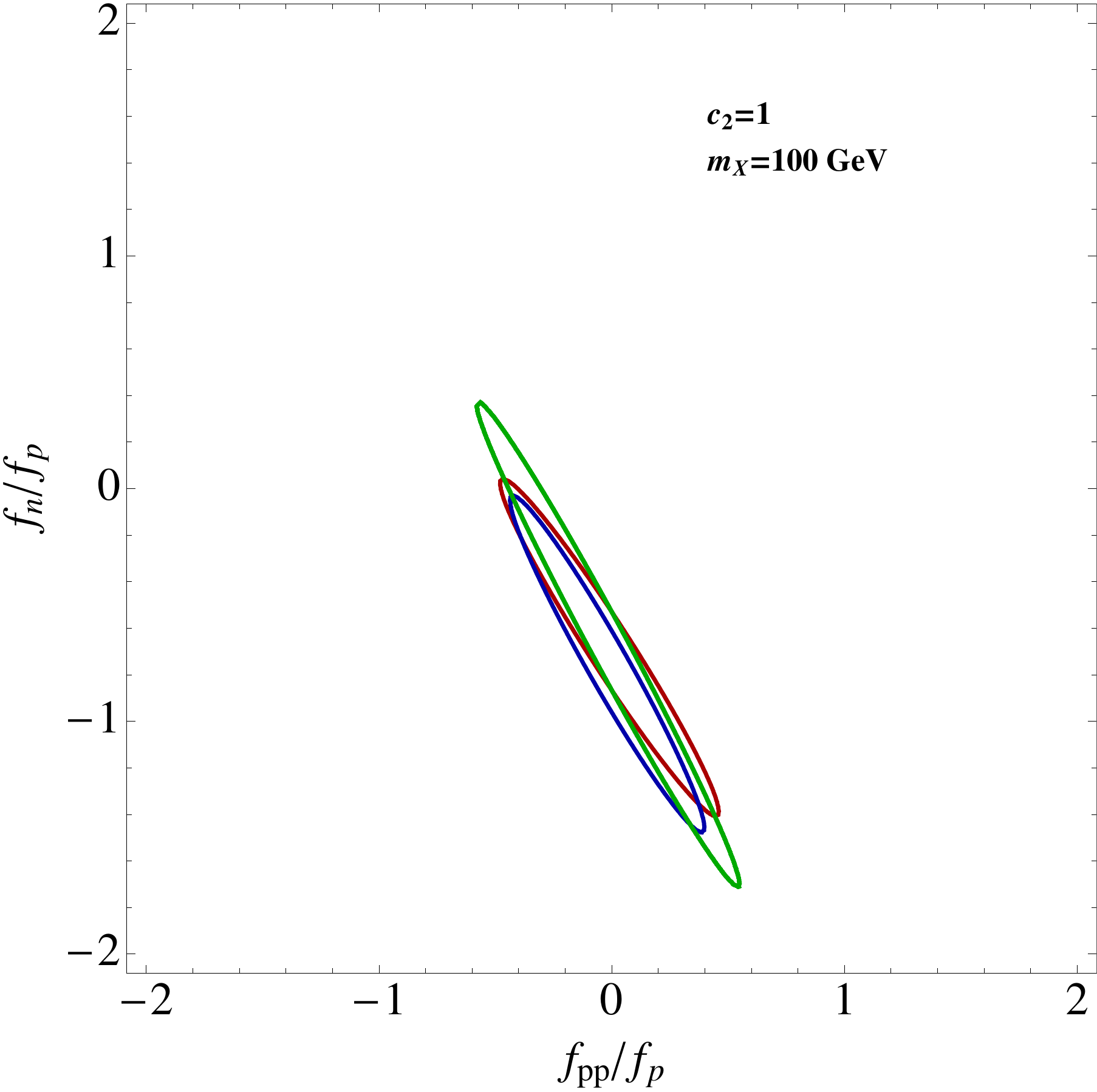}
\caption{Contour plots of $N_\Delta/N_\Delta^{\rm ref}=0.01$. The left column corresponds to $m_X=10\text{ GeV}$, the right column to $m_X=100\text{ GeV}$. The upper, middle, and lower plots are obtained with the phenomenological form factor $\overline F_{pp}^{\rm pheno}$ defined in (\ref{phenoFpp}) with $(c_1,c_2)=(0,-1)(0,0),(0,+1)$, respectively. Color coding: Red (Xenon100), Green (LUX), Blue (CDMS-Ge).}\label{fig:contourplots}
\end{center} 	
\end{figure}

We present our numerical results for three experiments: Xenon100,  LUX and CDMS-Ge in Figures \ref{fig:interplay}, \ref{fig:contourplots}. For the experimental details of the analysis we refer the reader to~\cite{Cirigliano:2013zta} and references therein. The color coding is Xenon100 (red), LUX (green), CDMS-Ge (blue). The impact of $f_{pp}$ is most conveniently discussed in terms of the ratio $N_\Delta/N^{\rm ref}_\Delta$, where $N^{\rm ref}_\Delta$ is defined with $f_{pp}=0, f_n=f_p$. This is sometimes referred to as the ``degradation plot" in the literature.

In Figure \ref{fig:interplay} we present degradation plots for $f_{pp}/f_{p}=-1$ (short dashed lines), $f_{pp}/f_{p}=0$ (solid lines) and $f_{pp}/f_{p}=1$ (long dashed lines). We consider the case of light DM, $m_X=10\,{\text{GeV}}$ (left column), and  $m_X=100\text{\,GeV}$ (right column). The top, center and bottom plots correspond to variations of the parameter $c_2$ (see our phenomenological two-body form factor) from $-1,0$, to $+1$.

For the case $f_{pp}/f_p=0$ we find the familiar LO degradation plot (see for instance~\cite{Feng:2013vaa} and references therein). By turning on non-zero values of $f_{pp}/f_{p}$, i.e. the two-body term, the entire plot is shifter and the minimum increases. This can be qualitatively understood looking at (\ref{CS}), from which we see that the minimum of the rate is always found at $f_{pp}=0$ ($x=0$), and gets displaced when $f_{pp}\neq0$.

Importantly, for $f_{pp}\neq0$ the Xenon100 and LUX experiments (and more generally other Xe-based detectors) observe different rates, despite having the same target nuclei. The reason is that the rate depends on the integrals in~(\ref{ND}), which are themselves functions of the energy threshold and the associated efficiency. If DM is detected, in principle this feature may be used to infer important information about the parameters $f_{pp}/f_p, f_n/f_p$. However, Fig. \ref{fig:interplay} suggests that the difference between the rates at different experiments significantly depends on the two-body form factor, so an accurate determination of $F^{(2)}$ would be required to draw any conclusion on $f_{pp}/f_p, f_n/f_p$.

To better appreciate this, in Figure \ref{fig:contourplots} we show contour lines of $N_\Delta/N^{\rm ref}_\Delta=0.01$ (similar contours are obtained for different values, as clear from Figure \ref{fig:interplay}) as a function of $f_{pp}/f_p, f_n/f_p$. As anticipated, we see that the amount of overlap among the (green, red, and blue) ellipses is critically sensitive to the two-body form factor.

In Figure \ref{fig:contourplots} we also observe a qualitative difference between light and heavy DM. For light DM the ellipses are very elongated, while for heavier masses they are more circular in shape. Again, this can be understood from (\ref{CS}). The astrophysical function in~(\ref{ND}) is more steeply falling for light DM, and this basically forces $E_R$ to acquire values close the lowest bin. In the approximation that the integrand in (\ref{ND}) is a delta function in $E_R$, the last term in (\ref{CS}) is very small and the domain $N_\Delta/N^{\rm ref}_\Delta={\rm const}$ is determined by a line $x\propto y$ with an experiment-dependent slope. For heavier DM the recoil spectrum has effectively a larger range and the second term in (\ref{CS}) becomes important.

The nature of a possible cancellation in the event rate is analogous to that invoked in isospin-violating DM (see e.g~\cite{Giuliani:2005my}\cite{Feng:2011vu}), as it arises from fine-tuning parameters that are naturally of the same order, but is qualitatively different for at least two reasons. First, the cancellation depends on $E_R$. Second, it involves two parameters --- i.e. $f_n/f_p, f_{pp}/f_p$ --- rather than one, so in principle it is possible to suppress $d\sigma_T/dE_{R}$ in two experiments simultaneously.

\section{The 2-nucleon form factor for $F_{\mu\nu}\widetilde{F}^{\mu\nu}$}
\label{sec:Ogammatilde}

To study DM scattering induced by $F_{\mu\nu}\widetilde{F}^{\mu\nu}$ one can proceed in complete analogy with the ${O}_{\gamma}$ operator: determine the non-relativistic amplitude for $NNX\to NNX$ and derive the corresponding nucleon potential. 

The crucial difference is that in the present case also neutrons contribute. The matrix element $\langle T_f|F_{\mu\nu}\widetilde{F}^{\mu\nu}|T_i\rangle$ is dominated by the {\emph{coherent}} scattering of one of the two photons on the proton charge, as in section~\ref{sec:2body}, and the {\emph{incoherent spin-dependent}} scattering of the second photon on the nucleon magnetic moment.

The multi-nucleon potential can again be written as in section~\ref{sec:2body}:
\begin{eqnarray}
&&\widetilde V_{ij}=\widetilde V_0\,\text{e}^{i{\bf q}\cdot {\bf R}}\, \widetilde{f}_{ij}({\bf q}, {\bf r}),
\end{eqnarray}
where now $\widetilde f_{ij}$ depends non-trivially on the nucleon spin. Up to $O({\bf q}^2/m_N^2)$ we find:
\ba
\widetilde{f}_{ij}({\bf q}, {\bf r})=i\int\frac{d^3 {\bf k}}{(2\pi)^3}\text{e}^{-i\bf{k}\cdot \bf{r}}\frac{\left[({\bf k}^2){\bf q}-({\bf k}\cdot{\bf q}){\bf k}\right]\cdot\overrightarrow\mu^+_{ij}+\left[({\bf q}^2){\bf k}-({\bf k}\cdot{\bf q}){\bf q}\right]\cdot\frac{\overrightarrow\mu^-_{ij}}{2}}{\left({\bf k}-\frac{\bf q}{2}\right)^2\left({\bf k}+\frac{\bf q}{2}\right)^2},
\ea
where we defined $\overrightarrow\mu^\pm_{ij}={\bf s}_i\mu^{\rm m}_ie_j\pm{\bf s}_j\mu^{\rm m}_je_i$, with ${\bf s}_i=\sigma/2$ the nucleon spin operator, and $\mu^{\rm m}_i, e_i$ the nucleon magnetic moment and electric charge. To keep our discussion general, we did not specify the DM bilinear coupling to $F_{\mu\nu}\widetilde{F}^{\mu\nu}$. This model-dependent contribution (generally momentum-dependent) is included in the coefficient $\widetilde V_0$. In other words, what we want to discuss here is the nuclear matrix element $\langle T_f|F_{\mu\nu}\widetilde{F}^{\mu\nu}|T_i\rangle$.

The nuclear form factor follows immediately from $\langle T_f|\sum_{i,j}\tilde V_{ij}|T_i\rangle$. One of the nucleon indices runs over the proton charge, and results in a single power of $Z$ in the amplitude. The remaining (spin-dependent) sum involves both neutrons and protons and is similar to that found in ordinary spin-dependent interactions.

Similarly to $O_\gamma$, the RG evolution of $\widetilde O_\gamma$ will induce contact (spin-dependent) interactions that reduce in the non-relativistic limit to the familiar 1-nucleon potential $\propto {\bf q}\cdot{\bf s}_i$. The ratio between long and short distance contributions to the amplitude for $XT\to XT$ scales as $ZQ_0/m_N$, as we found in section~\ref{sec:nucleonEFT}. An equivalent way to check this is to follow a logic similar to that of section~\ref{sec:nuclearEFT}, and add to the (spin-dependent) DM-target contact interaction a 1-loop diagram analogous to that evaluated in~\cite{Weiner:2012cb}, but now with the nuclear magnetic moment in one of the two $\gamma-T$ vertices.

\section{Realistic UV completions}
\label{sec:UV}

In this section we briefly comment on possible UV completions of $O_\gamma$ ($\widetilde O_\gamma$). We take $X$ to be a Majorana fermion for simplicity, but keep in mind that our results generalize to self-conjugate DM of any spin (baring naturalness issues in the case of scalar DM). Also, as emphasized in the introduction, generic UV completions of the DM polarizability operator will also contain $v^\mu v^\nu F_{\mu\alpha}F_{\nu\alpha}$~\footnote{With a coefficient suppressed by the ratio $m_X/m_*$ in the notation used in this Section.}. None of the results of this paper (nor of this Section) are affected by its presence.

\subsection{New physics at the weak scale}

The existing literature assumes that $O_{\gamma}$ is generated in isolation at some high scale $m_*$ by loops of some heavy mediator with electro-weak charges. However, generic field theories will also induce a lower dimensional coupling to the Higgs mass operator $H^\dagger H$ (and the quark mass operator). More precisely, on the basis of simple dimensional analysis we expect that when matching the UV completion at $m_*$ the effective Lagrangian will {\emph{generically}} contain --- in addition to the DM polarizability --- a contribution of order
\ba\label{CH}
\delta{\cal L}\supset C_H\,\frac{m_*^2}{\Lambda^3}\,\overline{X}{X}H^\dagger H, ~~~~~
C_H\sim \frac{\alpha}{4\pi}\,C_\gamma,
\ea
along with $C_q\sim\frac{\alpha}{4\pi}C_\gamma$. In terms of the fundamental constituents, (\ref{CH}) arises from diagrams similar to those leading to $O_\gamma$, where the $Z^0$ lines are closed in a loop and external $H$-legs are attached to it. We are here referring to a contribution at the matching scale $m_*$, not to an RG effect below $m_*$.

Now, while $C_H$ is suppressed compared to $C_\gamma$ at the cutoff, the latter contributes in DD at one-loop order. A back of the envelop calculation tells us that the ratio between the DM-nucleon cross section induced by $O_{\gamma}$ and that of $O_H$ is parametrically suppressed by $\left({m_h}/{m_*}\right)^4$. Numerically, the coupling $O_{\gamma}$ will be relevant (i.e. $f_{pp}/f_p\sim1$) only if the electro-weak charged mediators have masses $m_*$ below a few hundred GeV. This holds irrespective of the actual magnitude of the signal, and is a priori independent from the usual WIMP miracle and arguments based on the naturalness of the SM: generic models in which $f_{pp}/f_p\sim1$ will have new physics accessible at colliders. The strongest constraint on the messenger mass scale in these scenarios come from direct searches at LEP and the LHC, and for relatively short lived particles are compatible with new physics at the weak scale.~\footnote{For a broad perspective on the current bounds, see for instance \cite{Aad:2014vma}\cite{Khachatryan:2014qwa}. An explicit analysis of some model relevant to our discussion was presented in~\cite{Weiner:2012gm}\cite{Liu:2013gba}. The results of these papers are relevant to our scenario, even though in some of those models $O_\gamma$ is not relevant to DD experiments.}

The situation is unchanged in UV-complete models for $\widetilde O_\gamma$, since $\overline{X}{X}H^\dagger H$ is allowed by all relevant symmetries.

It should be clear that, in analogy with the more familiar hierarchy problem, one can evade the natural expectation (\ref{CH}) by fine-tuning, or invoking Supersymmetry and/or Higgs compositeness at scales $\ll m_*$. In this sense we claim that a UV completion for (\ref{Ogamma}) at the weak scale is {{generic}}, but not strictly necessary.

\subsection{Large DD rates and suppressed indirect signatures}

Much can be learnt about possible UV completions by making a rough estimate of the DD rate generated by $O_\gamma$. 

First of all, in a healthy theory (\ref{Ogamma}) will be generated at one-loop level  
\ba\label{Cgamma}
\frac{C_\gamma}{\Lambda^3}=e^2\frac{g_*^2}{16\pi^2}\frac{m_X}{m_*^4},
\ea
with $g_*$ a typical coupling (the power of $m_X$ is expected if $m_X\ll m_*$). Using the results of section~\ref{sec:nucleonEFT} we estimate a nucleon-DM cross section (on Xe) of order
\ba\label{rateNX}
\sigma_{N}&\sim&8\left(\frac{\alpha}{\pi}\right)^2\left(\mu_NQ_0\frac{C_\gamma}{\Lambda^3}\right)^2\frac{Z^4}{A^2}\\\no
&\sim&6\times10^{-46}~{\rm cm^2}~\left(\frac{g_*}{4\pi}\right)^4\left(\frac{100~{\rm GeV}}{m_*}\right)^8\left(\frac{m_X}{100~{\rm GeV}}\right)^2~~~~~~({\rm for~Xe})
\ea
with $\mu_N$ the reduced DM-nucleon mass, and $1/Q_0\sim A^{1/3}$ fm a measure of the radius of the target nuclei. Together with the collider bounds mentioned at the end of the previous section, this estimate strongly disfavors DM detection in forthcoming DD experiments. This is even more true for $\widetilde O_\gamma$, which has a much lower rate.

However, this estimate fails if the heavy messengers interact with DM primarily via a light scalar. Consider first $O_\gamma$ and postulate the dark sector couples to a light dilaton $\phi$ of mass $m_\phi\ll m_X, m_*$. In this case the high energy theory contains a (loop-order) coupling $F_{\mu\nu}^2 \phi$ and, provided the trilinear $XX\phi$ is unsuppressed (typically of order $g_* m_X/m_*$), $O_\gamma$ will be dominated by the tree-level exchange of $\phi$. For $\phi$ heavier than $\sim100$ MeV the rate (\ref{rateNX}) receives an enhancement of $\sim({m_*^2}/{m_\phi^2})^2$. Now nucleon cross sections of order $10^{-44}~{\rm cm}^2$ become plausible for $m_X,m_*\sim100$ GeV provided $m_\phi/g_*$ is in the few GeV range. Still, natural considerations suggest $m_\phi^2\gtrsim e^2m_*^2/16\pi^2$, from which follows that $\sigma_N$s significantly larger than a few times $10^{-44}~{\rm cm}^2(100~{\rm GeV}/m_*)^6$ may be interpreted as indirect signature of an unnaturally light scalar.

It is hard to imagine a mechanism to enhance the rate of $\widetilde O_\gamma$ up to similar values. Even if we add an axion $a$ with mass $m^2_a\ll q^2$ the cross section will be typically too small to be detected. The reason is that the power of $1/q^2$ from the propagator is compensated by a factor ${\bf q}\cdot{\bf s}_{\rm DM}$ from the axion-DM coupling, and ${\bf q}\cdot{\bf s}_{T}$ from the axion-target coupling. Yet, axion-mediated DM interactions can naturally suppress the ``dangerous" DM-Higgs coupling of the previous subsection, and typically have $f_{pp}/f_p\sim1$.

An obvious question is whether indirect gamma-ray signatures of, say $O_\gamma$, can be relevant. Importantly, for $m_\phi\ll m_X$ the s-channel exchange of $\phi$ will lead to:
\ba\label{annih}
\langle\sigma_{\gamma\gamma}v\rangle\sim v^2\frac{e^4}{\pi}\left(\frac{g_*^2}{16\pi^2}\right)^2\frac{m_X^2}{m_*^4}.
\ea
Because of the much larger momentum flowing in the $\phi$ propagator, the result is a factor $\sim(m_\phi/m_X)^4$ {\emph{smaller}} than that obtained assuming that the same scale suppressing $O_\gamma$ in DD experiments also controls indirect signals. Indirect detection rates close to the current sensitivities are still possible (see for instance~\cite{Fedderke:2013pbc}), especially for scalar DM that does not suffer from the velocity suppression in~(\ref{annih}).

\section{Conclusions}
\label{conclusions}

We presented a detailed analysis of the DD signature induced by the DM polarizability operator $O_\gamma$, see eq.(\ref{Ogamma}), and derived the associated nuclear form factor as a function of the 2-nucleon density. The relevance of 2-body densities was previously pointed out in~\cite{Prezeau:2003sv}\cite{Cirigliano:2012pq}, but in those cases the effect was subleading in the chiral expansion and for generic choices of the parameters. To the best of our knowledge, $O_\gamma$ provides the first known example of DM scattering {\emph{dominated}} by a 2-body nuclear form factor. The long range force mediated by the photon is key to our result.

The multi-body nature of the interaction implies that the scattering rate varies significantly from experiment to experiment, and that DD experiments with the same target nuclei are expected to measure different rates. The presence of destructive interference in a non-negligible portion of the parameter space makes these scenarios especially interesting in light of current anomalies.

DM scattering via $F_{\mu\nu}\widetilde F^{\mu\nu}$ was also discussed. The novel feature here is that the 2-body interaction actually describes an example of {\emph{coherent}} and simultaneously {\emph{spin-dependent}} DM scattering.

One expects $F_{\mu\nu}F^{\mu\nu}$ and $F_{\mu\nu}\widetilde F^{\mu\nu}$ to be important to DD experiments in a large class of models with self-conjugate DM. We showed that realistic scenarios in which these operators are relevant to DM detection have new physics at scales accessible at colliders. Furthermore, large DD rates are indirect evidence of exotic light scalars, and are typically accompanied by suppressed gamma-ray signatures.

\acknowledgments

We would like to thank John F. Donoghue for useful conversations. The work of LV was supported in part by the NSF Grant No. PHY-0968854, No. PHY-0910467, by the Maryland Center for Fundamental Physics, and by the National Science Foundation under Grant No. PHY11-25915.

\appendix

\section{Comparison with previous work}
\label{sec:vanish}

It is instructive to compare our approach to an EFT of the nucleus. There are several diagrams contributing to $XT\to XT$ in this latter formalism, and their sum should reproduce our result (\ref{TV}). In particular, one contribution comes from a 1-loop process analogous to the left diagram in fig~\ref{fig:FD}, with the external lines understood as ground state nucleus whereas the internal solid line as all possible virtual states allowed by the symmetries. We would like to show that these diagrams correctly reproduce the structure of (\ref{TV})-(\ref{MNR}), as expected.

To see this note that the 1-loop diagrams can be formally written as
\ba
\int\frac{d^4l}{(2\pi)^4}
\frac{{\bf l}^2-{\bf q}^2/4}
{\left[\left(l+\frac{q}{2}\right)^2+i\epsilon\right]\left[\left({l}-\frac{q}{2}\right)^2+i\epsilon\right]}
\langle T_f |J^0 G J^0|T_i\rangle,
\ea
where $G=1/(E-H_{\rm strong}+i\epsilon)$ the multi-nucleon propagator including all possible insertions of the strong interactions, $E$ the energy of the intermediate state, and $J^0$ the zeroth component of the proton electro-magnetic current. Plugging in a complete set, the above expression becomes
\ba\label{A1}
\int\frac{d^4l}{(2\pi)^4}
\frac{{\bf l}^2-{\bf q}^2/4}
{\left[\left(l+\frac{q}{2}\right)^2+i\epsilon\right]\left[\left({l}-\frac{q}{2}\right)^2+i\epsilon\right]}
\sum_{n}\frac{\langle T_f |J^0\left(l+\frac{q}{2}\right)|n\rangle  \langle n| J^0\left(l-\frac{q}{2}\right)|T_i\rangle}{l^0-\Delta m_n+i\epsilon},
\ea
where $\Delta m_{n}$ is the difference between the mass of $|n\rangle$ and the nuclear ground state. Intermediate states with $\Delta m_{n}\gtrsim Q_0\sim100$ MeV decouple and are not relevant. On the other hand, those with $\Delta m_{n}\ll Q_0$ cannot be neglected. One can verify that the integral in $l^0$ is dominated by scales of order $Q_0$. Therefore, replacing $l^0-\Delta m_n$ with $l^0$ results in a small ${\cal O}(\Delta m_n/Q_0)$ error. Ignoring the latter, we find that the sum in (\ref{A1}) reduces to $\frac{1}{l^0+i\epsilon}\langle T_f |J^0 J^0|T_i\rangle$. Performing the integral in $l^0$ using the residue theorem we obtain an expression completely analogous to (\ref{MNR}), as promised.

The latter diagram has been considered by Weiner and Yavin for $|n\rangle=|T\rangle$~\cite{Weiner:2012cb}. Our expression (\ref{MNR}) reduces to their result if we ignore proton-proton correlations, i.e. let $F^{(2)}\to F^{(1)}_pF^{(1)}_p$, which effectively parametrizes the contribution of the virtual excited states $|n\rangle\neq|T\rangle$. Indeed it is readily seen that (\ref{A1}) formally becomes eq.(A7) of~\cite{Weiner:2012cb} when $|n\rangle\langle n|=1$ and $Z\gg1$.~\footnote{To see this more explicitly one should redefine the velocity of the target in order to remove the redundant variable $\tilde p$, and neglect $q^0\sim {\bf q}^2/2m_T\ll {\bf q}^2/Q_0$ in eq.(A7) of~\cite{Weiner:2012cb}.}

Direct detection mediated by $O_\gamma$ was also discussed in~\cite{Frandsen:2012db} and~\cite{Crivellin:2014gpa}. There the RG evolution down to $\mu\sim m_c$ was studied in some detail, and an approximate expression for the DM couplings to nucleons was derived. However, our discussion in section~\ref{sec:nucleon} suggests that this last step is plagued by $O(1)$ uncertainties arising from the RG evolution down to $\mu\sim m_N$, which forced us to treat $C_{p,n}(m_N)$ as ``free" parameters. As in~\cite{Weiner:2012cb}, the authors of~\cite{Frandsen:2012db}\cite{Crivellin:2014gpa} did not include proton-proton correlations. In particular, when $c_i=0$ the form factor $\overline F_{pp}^{\rm pheno}$ defined in eq.(\ref{phenoFpp}) reduces to the quantity $F_{\rm Ray}$ introduced in~\cite{Frandsen:2012db}. In general we find:
\ba
\overline F_{pp}^{\rm pheno}=\left[1+\frac{1}{4}c_1\left(1-\bar q\frac{d}{d\bar q}-\bar q^2\right)+c_2\bar q^2\right]F_{\rm Ray}(\bar q),
\ea
where
\ba\no
F_{\rm Ray}(\bar{q})=1-\frac{\pi ^{3/2}}{2 \sqrt{2}}{\bar{q}}+\frac{7}{6} {\bar{q}}^2-\frac{71}{360} {\bar{q}}^4+\frac{319}{8400} {\bar{q}}^6-\frac{5419}{846720} {\bar{q}}^8+\frac{22369}{23950080} {\bar{q}}^{10}+\mathcal{O}(\bar{q}^{12}).
\ea
An expansion for $\overline F_{pp}^{\rm pheno}$ up to $O(q^2)$ is given in~(\ref{phenoFpp})

\end{document}